\documentclass{article}
\usepackage{spconf,amsmath,graphicx,hyperref,booktabs}

\usepackage{wasysym}

\usepackage{textcomp}  % Required for encoding \textbigcircle
\usepackage{scalerel}  % Required for emoji \scalerel
\usepackage{amssymb} % For \mathbb

\newcommand{\emoji}[1]{\scalerel*{\includegraphics{imgs/emojis/#1.png}}{\textrm{\textbigcircle}}}

\title{VIDEO-TO-MUSIC GENERATION FOR GAMEPLAY VIDEOS}
\name{Felipe Marra\textsuperscript{{\quarternote}}, Lucas N. Ferreira\textsuperscript{\twonotes}}
\twoauthors
  {Felipe Marra}
	{Universidade Federal de Viçosa \\
	Departamento de Informática \\
    Viçosa, Minas Gerais, Brazil}
 {Lucas N. Ferreira}
    {Universidade Federal de Minas Gerais \\
    Departamento de Ciência da Computação \\
    Belo Horizonte, Minas Gerais, Brazil}
\begin{document}
%\ninept
%
\maketitle
%

% história
% motivação -> dados sintéticos que não estão cobertos em datasets anteriores
% modelo -> ablation, baselines & listening test

% Abstract
% -> ninguém tá fazendo com rendered videos and synthetic audios. foco no ser o primeiro e ninguém tá fazendo
% -> Furthermore -> more data, genre, out of distribution data
% explicar que construímos um modelo SIMLPES, fizemos ablation de modalidades diferentes -> somos o primeiro em video games
% comparamos com resultados de forma objetiva, fizemos o listening test e o resultado é tal
% Tirar o foco dos modelos (ViViT, ViT, T5) e focar nas modalidades

\begin{abstract}
Video-to-music models have advanced considerably in the last few years, particularly in film and music video applications. In this paper, we investigate this problem in the video game domain, which introduces new challenges for these models: video frames are rendered graphics, music is mostly synthetic audio, and soundtracks loop across entire levels rather than following on-screen events. We introduce a new dataset of 217.6 hours of Super Nintendo (SNES) gameplay video paired with 485 hours of clean soundtracks, free of sound effects and voice-overs, matched to gameplay audio via audio fingerprinting. With this dataset, we train a simple encoder-decoder transformer that passes video features directly to a MusicGen decoder, comparing different encoding strategies: textual descriptions (T5), independent frames (ViT), or spatiotemporal patches (ViViT). Each encoder is tested both frozen and fine-tuned, while the decoder is always fine-tuned. Frozen encoders match or outperform their fine-tuned counterparts on every metric, and the frozen ViViT achieves the best overall results. We compare this model with state-of-the-art baselines using both objective metrics and a listening study ($N = 96$). Despite having up to 18\% fewer parameters, our model outperforms all baselines on objective metrics, surpasses GVMGen in the listening study, and performs comparably to OSSL.
\end{abstract}

% INDEX TERMS (KEYWORDS): Enter up to 5 keywords separated by commas.
\begin{keywords}
multimedia, video-to-music, video games.
\end{keywords}
%

% Intro
% foco é resolver o problema de videogames
% pode explicar as diferenças de timbre, mas depois de especificar o lance dos dados sintéticos como no ablation
% Since this is the first... we propose a simple model... ablation with encoders of different modalities (doesn't matter the encoder itself) tuned and frozen...
% 

\section{Introduction}
\label{sec:intro}

Video-to-music (V2M) research primarily focuses on film~\cite{ossl, vidmuse} and music videos~\cite{mumu-llama,sym_mv,v2meow}, domains with real footage where the generated music needs to be temporally aligned with the video. Video games are another promising application: V2M models could help independent developers, who often lack the resources for custom soundtracks, create music that fits their game's visuals. However, video games pose a different challenge: both graphics and audio are synthetic, and soundtracks often loop across entire levels rather than following on-screen events. It remains unclear whether V2M models, built on encoders pre-trained on real-world data, transfer to this domain.

We extend V2M to retro video games, a style that remains popular among independent titles. We introduce a Super Nintendo (SNES) dataset that pairs gameplay videos with clean soundtracks, free of voice-overs and sound effects. By fingerprinting 10-second YouTube clips against the original soundtracks and downsampling across 11 genres for balance, we obtain 217.6 hours of video paired with 485 hours of music\footnote{https://felipemarra.github.io/demo-v2m-4-gameplay-videos-v1/}.

With this dataset, we train a simple encoder-decoder transformer that passes video features directly to a pre-trained MusicGen~\cite{musicgen} decoder via cross-attention. To find the best video encoder for this domain, we compare three approaches common in the V2M literature: describing the video textually (T5)~\cite{herrmann1}, encoding it frame by frame (ViT)~\cite{gvmgen}, or fusing frames across time (ViViT)~\cite{ossl}. Each encoder is initialized with pre-trained weights and tested both frozen and fine-tuned, while the decoder is always fine-tuned. Frozen encoders match or outperform their fine-tuned counterparts on every metric, and the frozen ViViT achieves the best overall results, with the lowest FAD~\cite{fad} and competitive KL divergence and ImageBind~\cite{imagebind} scores.

We compare this model with state-of-the-art (SOTA) baselines: Herrmann-1~\cite{herrmann1}, OSSL~\cite{ossl}, and GVMGen~\cite{gvmgen}. Our model outperforms all baselines on objective metrics. In a listening study ($N=96$), it outperforms GVMGen and performs comparably to OSSL, despite having up to 18\% fewer parameters. These results suggest that frozen Transformer encoders pre-trained on real-world data can generalize to out-of-distribution domains such as retro video games. 
% Our code and dataset (video URLs and clean soundtrack mappings) are available\footnote{https://felipemarra.github.io/demo-v2m-4-gameplay-videos-v1/}.

% within constrained domains like retro games, simpler Transformer architectures can generalize to out-of-distribution data and exceed specialized V2M models.

% strict timbral and polyphonic constraints. This lower audio variance, facilitates reliable evaluation. It also allows us to test how visual encoders pre-trained on real footage transfer to out-of-distribution domains.

% #################################
% Related Work
% #################################

% Move the GVMGen argument against ViViT to the conclusion (or remove it)

% direct mapping -> Apesar do foco ser áudio, existe em MIDI, NES-MVDB cria um dataset parecido com o nosso com CMT, que tem foco em MIDI
% No último parágrafo, que é feito pra apresentar o contraste, já falar do source separation e de outros args da sec 3.
% Se não encaixar NES-MVDB no fim da sec 2, pode fazer sentido mover pra intro

\section{Related Work}
\label{sec:related_work}

\textbf{Video-to-Music Models.} V2M generation conditions music on a sequence of video frames. Given its proximity to text-to-music (T2M) generation, V2M models largely build on pre-trained T2M models, which generate music from textual descriptions. 
% These models synthesize audio directly by leveraging neural audio codecs~\cite{improved_rvqgan} that compress waveforms into compact discrete tokens ~\cite{musiclm, stable_audio, musicgen}. 
% MusicLM~\cite{musiclm} and Stable Audio~\cite{stable_audio} condition on MuLan~\cite{mulan} and CLAP~\cite{microsoft_clap} embeddings, respectively, while 
MusicGen~\cite{musicgen} is one of the most widely T2M models adopted as a base for V2M generation. It leverages an neural audio codec~\cite{encodec} to compress waveforms into compact discrete tokens, and predicts interleaved codebook streams in parallel, conditioned on T5~\cite{t5} embeddings. Given this relation to T2M generation, we categorize V2M models by how they use text to bridge visual and musical modalities: \textit{text as interface}, \textit{text embedding as interface}, and \textit{direct mapping}.

\textit{Text as Interface.} These approaches extract visual features, convert them into a musical description, often via an LLM, and pass that description to a T2M generator. Because these components can remain frozen, this paradigm can use unpaired video and music data. Herrmann-1~\cite{herrmann1} tackles the film domain by transforming scene descriptions, detected emotions, and speech into music prompts for MusicGen. CoT-VTM~\cite{cot_vtm} is similar, but distills the pipeline into a single model that outputs directly into the T5 embedding space. SONIQUE~\cite{sonique} extracts visual tags to condition a pre-trained Stable Audio model.

\textit{Text Embedding as Interface.} These methods map visual features into the text embedding space of a music decoder via an adapter, offering greater representational flexibility than discrete text. OSSL~\cite{ossl} uses a LoRA adapter to bridge a frozen ViViT~\cite{vivit} video encoder with the T5 text encoder of MusicGen. MuMu-LLaMA~\cite{mumu-llama} employs modality-specific adapters to build a multimodal LLM whose outputs are mapped to T5 tokens for synthesis. VidMusician~\cite{vidmusician} extends this paradigm with time alignment, using CLIP-based [CLS] tokens for semantic conditioning and high-frequency patch-level tokens for rhythmic synchronization.

\textit{Direct Mapping.} Bypassing text entirely, these models map visual features directly to the music decoder. This typically requires training the decoder, as visual features lie outside the text encoder's distribution, and suits tasks that require temporal alignment. GVMGen~\cite{gvmgen} uses a frozen ViT~\cite{vit} with a trainable Transformer adapter, arguing that spatiotemporal fusion, as in ViViT, degrades frame-level alignment. VidMuse~\cite{vidmuse} employs an LSTM adapter with a sliding window to provide both global guidance and local adaptation. V2Meow~\cite{v2meow} extracts features via 3D-CNN, CLIP, and ViT-VQGAN, training a MusicLM-based decoder on a large-scale music video dataset filtered from YouTube-8M~\cite{youtube8m}.

\textbf{Video Game Music.} Music generation for video games has mostly been studied in the symbolic domain. NES-MDB~\cite{nes_mdb} provides multi-instrumental NES soundtracks with expressive performance attributes, and LakhNES~\cite{lakhnes} generates NES-style music with a Transformer pre-trained on the Lakh MIDI dataset and fine-tuned on NES-MDB, without conditioning on gameplay. Closest to our work, NES-MVDB~\cite{nes_vmdb} pairs NES gameplay videos with symbolic music, using rule-based video features for MIDI generation. In contrast, we generate audio and target the SNES, whose 8-channel sample-based audio is considerably richer than the NES's 5-channel synthesized sound and whose library contains more games.

\section{SNES Video-to-Music Dataset}
\label{sec:dataset}

\textbf{Collection \& Pre-Processing.} We started by scraping the original SNES soundtracks from Zophar's Domain,\footnote{\url{https://www.zophar.net/music/nintendo-snes-spc}} an online repository of crowd-sourced video game assets. Following~\cite{nes_vmdb}, we filtered out all audio files shorter than 8 seconds and, for each of the 1,604 resulting games, queried YouTube for a gameplay video using ``\emph{\{GAME\_NAME\} Nintendo SNES Longplay}'', where \emph{\{GAME\_NAME\}} is the game's title on Zophar's Domain, keeping the top result. We partitioned the videos into 10-second clips to match the ViViT encoder's input while minimizing the likelihood of capturing track transitions within a single clip.

\textbf{Noisy to Clean Soundtracks.} To map noisy gameplay audio to the clean original soundtracks from Zophar's Domain, we built game-specific fingerprint databases using Dejavu~\cite{dejavu}, which ranks candidate tracks by a confidence score, defined as the fraction of successful fingerprint matches. We queried each clip's audio against its game's database and assigned it the top-ranked clean track if its confidence score was above $0.01$, a value chosen empirically. To evaluate the quality of this mapping, we manually annotated 10 hours of video, balanced by genre (approximately 0.9 hours per genre), with the start and end timestamps of each music track. Compared against these annotations, the assigned tracks achieved an average accuracy of 77.36\%.

\textbf{Downsampling.} To prevent genre dominance, we mapped titles to an 11-genre taxonomy~\cite{nes_vmdb}: Action, Adventure, Fighting, Platform, Puzzle, RPG, Racing, Shooter, Simulation, Sports, and Strategy. We identified each title's genre by querying its name on Wikidata and using an LLM (DeepSeek R1-70B) to map the retrieved metadata to the most appropriate genre. We then downsampled clips from majority genres to balance them with the minority genres, while preserving at least one clip of every unique background track to ensure diversity. The final dataset comprises 78,344 clips, totaling 217.6 hours of video, paired with 16,808 unique background tracks, totaling 485 hours of clean audio.
\section{Video Encoders}
\label{sec:video_encoders}

% We built the model, ablation and etc. (remove the compare as if we did not built it)
% Best encodes video game metrics

With our SNES dataset, we compare three of the most common encoders in autoregressive V2M approaches: T5~\cite{t5}, ViT~\cite{vit}, and ViViT~\cite{vivit}. We build three encoder-decoder models, one per encoder, while keeping the same MusicGen decoder in all of them. In each model, the encoder output is linearly projected to the decoder's dimension and passed to the decoder via cross-attention.

\textbf{T5.} This model follows a text-as-interface approach, where visual information is translated into text before generation. We use VideoLLaMA 3~\cite{video_llama3} to generate detailed captions for each 10-second clip, with a prompt that asks the model to describe interface options and backgrounds for menu scenes, and actions, environments, movement speed, and mechanics for gameplay. A T5 encoder then maps each caption to a sequence of contextual embeddings $E = (e_1, \dots, e_k)$, where $k$ is the number of T5 tokens. Notably, we omit the intermediate step of using an LLM to ``translate'' scene descriptions into musical prompts, forcing T5 to represent the raw visual semantics and allowing a direct comparison with the purely visual encoders.

\textbf{ViT.} The remaining two models follow a direct mapping approach. The ViT-based model represents a video as a temporal sequence of independent frames. From each clip, we sample 32 evenly spaced frames $F = (f_1, \dots, f_{32})$, resize the short size to $224$ and center crop to $224 \times 224$, and encode each one with ViT-B/16, producing 196 patch embeddings and one prepended [CLS] token. Since the [CLS] token summarizes the frame, we keep only the sequence of [CLS] vectors $C = (c_1, \dots, c_{32})$, yielding a compact frame-level temporal representation.

\textbf{ViViT.} The ViViT-based model uses the same 32 frames, but projects them into non-overlapping $2 \times 16 \times 16$ tubelets spanning the spatiotemporal volume, capturing spatial and temporal information jointly in a single projection. This produces a sequence of spatiotemporal embeddings $V = (v_1, \dots, v_n)$, with $n = 3{,}136$, where each token represents a local region of the gameplay. Unlike in ViT, we omit the global [CLS] token to preserve local spatiotemporal detail rather than a single clip-level summary.
% #################################
% Experiments
% #################################
\section{Experiments and Results}

% Subsec exp setup
% Subsec ablation -> Ver meu comentário sobre essa ordem
% Objective metrics
% Listening test

% #################################
% Subsec: Experimental Setup.
% #################################

% We chose bigger validation because

\textbf{Experimental Setup.} We initialized each encoder with pre-trained weights: ViViT-B~\cite{vivit}  on Kinetics-400 action classification, ViT-B~\cite{vit} on ImageNet classification, and T5-base~\cite{t5} on the Colossal Clean Crawled Corpus. Each encoder was trained in two regimes: frozen and fully fine-tuned on our dataset. The decoder is initialized with pre-trained MusicGen medium weights and is always fine-tuned. The projection layer is always randomly initialized. We split the dataset by genre into 50\% for training, 40\% for validation, and 10\% for testing; the large validation split is used to fine-tune the ImageBind~\cite{imagebind} model used in our objective metrics. Following~\cite{musicgen}, we trained all models for 75 epochs of 2,000 updates each, with a batch size of 6 as in GVMGen~\cite{gvmgen}, using AdamW with a learning rate of $10^{-4}$ and a cosine scheduler. Training used Distributed Data Parallel across three NVIDIA A100 GPUs (80 GB). At inference, we generate 30-second audio at 32 kHz using top-$k$ sampling with $k=250$ and a classifier-free guidance scale of 3.

% #################################
% Subsec: Objective Metrics
% #################################

\textbf{Objective Metrics.} Following standard V2M practice~\cite{gvmgen, ossl, vidmuse}, we report $\text{FAD}_{\text{VGGish}}$~\cite{fad, vggish} for audio quality, $\text{KLD}_{\text{PaSST}}$~\cite{passt} for semantic similarity to the reference, and the ImageBind~\cite{imagebind} (IB) score for video-music alignment. To compute these metrics, each model generates one music piece for every clip in the test split, then scores are calculated and averaged per genre. FAD and KLD compare each generated piece with the clip's mapped ground-truth soundtrack (Section~\ref{sec:dataset}); IB is the cosine similarity between the embeddings of each clip and its generated piece. Since IB was not trained on SNES music, we fine-tuned its audio and video encoders contrastively on our validation split, following a linear-probing-then-fine-tuning strategy~\cite{finetuning}: 5 epochs tuning only the last layer, followed by 5 epochs of LoRA with the last layer frozen. We denote this tuned version as $\text{IB}_{\text{tuned}}$. The test split is never used for tuning.

% #################################
% Subsec: Ablation
% #################################

\textbf{Encoders Comparison.} Table~\ref{tab:objective} reports the mean and standard deviation across genres of $\text{FAD}_{\text{VGGish}}$, $\text{KLD}_{\text{PaSST}}$, and $\text{IB}_{\text{tuned}}$ for the T5, ViT, and ViViT encoders, with fine-tuned ones denoted by~\emoji{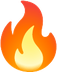} and frozen ones by~\emoji{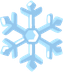}. We compare all model variants against the original MusicGen model (\emoji{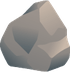} MGen)~\cite{musicgen}, which is not fine-tuned on our dataset and is conditioned on the same captions as our T5 model. All \emoji{snowflake} variants outperform it on every metric, confirming the benefit of training the MusicGen decoder on our dataset. For every metric, \emoji{snowflake} models perform better than or equal to their \emoji{fire} counterparts. We hypothesize that tuning only the decoder lets the cross-attention adapt to stable pre-trained features, whereas tuning both at once distorts these features~\cite{finetuning}. KLD differences among \emoji{snowflake} models are subtle, with T5, ViViT, and ViT scoring $0.71\pm0.68$, $0.72\pm0.60$, and $0.73\pm0.63$, respectively. IB scores are also similar across models, with \emoji{snowflake} ViT and ViViT tied for the highest mean and lowest variance ($0.49\pm0.02$), suggesting that all encoders extract useful semantic information for music generation. In FAD, \emoji{snowflake} ViViT achieves the lowest mean and variance ($10.53\pm3.60$), making it the only model ranked in the top two on all metrics.

% While the GVMGen~\cite{gvmgen} authors argue that temporal fusion is unsuitable for time-aligned domains, our results suggest it may benefit domains where tight temporal alignment is less important, such as video games.

%The~\emoji{snowflake} ViViT encoder achieved the best overall performance, yielding the lowest FAD ($10.53$) and highest IB ($0.49$) scores.
%In KLD, it performs comparably to~\emoji{snowflake} T5:~\emoji{snowflake} T5 achieved $0.71$ and ~\emoji{snowflake} ViViT $0.72$. The frame-based ViT performed strongly on the other metrics ($FAD=10.57$ and $IB=0.49$), suggesting that high-resolution spatial features are highly informative for the SNES domain. In contrast, the worst metrics were given by the tuned encoders, with $0.927$ of KLD for~\emoji{fire} ViT, which also shows high variance across genres, $11.575$ of FAD and $0.467$ of IB for~\emoji{fire} ViViT. %This indicates that tuning only the encoder will allow the cross-attention to adapt to the new encoder, while tuning both at the same time degrades the cross-attention features. 
%The superior performance of frozen encoders across all architectures suggests that foundational pre-trained hierarchies transfer to retro-game aesthetics, whereas domain-specific fine-tuning on a smaller corpus leads to representation drift. Since~\emoji{snowflake} ViViT is our best model, in Figure~\ref{fig:heatmap} we compare it with the SOTA approaches. 

% #################################
% Table: Ablation
% #################################

\begin{table}[t]
\centering
\setlength{\tabcolsep}{3pt}
% \resizebox{\columnwidth}{!}{
\begin{tabular}{lccc}

\toprule

& $\mathbf{KLD_{PaSST}~\downarrow}$ & $\mathbf{FAD_{VGGish}~\downarrow}$ & $\mathbf{IB_{tuned}~\uparrow}$  \\

\midrule

% TODO
% \textbf{Human} & $0.00$ & $8.07$ & $0.29$ & - \\

\textbf{\emoji{stone} MGen} & $0.85\pm0.64$ & $13.90\pm4.09$ & $0.45\pm0.02$ \\

\textbf{\emoji{fire}T5} & $0.80\pm0.66$ & $11.32\pm4.06$ & $0.47\pm0.02$ \\

\textbf{\emoji{snowflake}T5} & $\mathbf{0.71\pm0.68}$ & $11.07\pm4.27$ & $0.47\pm0.02$ \\

\textbf{\emoji{fire}ViT} & $0.93\pm0.98$ & $10.98\pm3.95$ & $\mathbf{0.49\pm0.05}$ \\

\textbf{\emoji{snowflake}ViT} & $0.73\pm0.63$ & $10.57\pm3.97$ & $\mathbf{0.49\pm0.02}$ \\

\textbf{\emoji{fire}ViViT} & $0.76\pm0.55$ & $11.57\pm4.09$ & $0.47\pm0.03$ \\

\textbf{\emoji{snowflake}ViViT} & $0.72\pm0.60$ & $\mathbf{10.53\pm3.60}$ & $\mathbf{0.49\pm0.02}$ \\

% \midrule
% \textbf{Herrmann-1} & $0.892\pm0.581$ & $115.477\pm4.573$ & $0.440\pm0.029$ & $2.22$ \\ 
% \textbf{GVMGen} & $0.860\pm0.602$ & $17.226\pm4.893$ & $0.425\pm0.063$ & $2.22$ \\ 
% \textbf{OSSL} & $0.814\pm0.691$ & $12.064\pm3.845$ & $0.433\pm0.020$ & $2.35$ \\
% \textbf{\emoji{fire}GVMGen} & $0.804\pm0.593$ & $12.760\pm4.215$ & $0.431\pm0.029$ & $2.22$ \\ 
% \textbf{\emoji{fire}OSSL} & $0.819\pm0.706$ & $11.696\pm3.973$ & $0.435\pm0.017$ & $2.35$ \\ 
\bottomrule
\end{tabular}
% }
\caption{Encoders comparison. Metrics are averaged across genres.~\protect\emoji{fire} denotes a fine-tuned encoder while~\protect\emoji{snowflake} denotes a frozen pre-trained one.~\protect\emoji{stone} MGen is MusicGen Base.}
\label{tab:objective}
\end{table}

% #################################
% Subsec: SOTA Comparison
% #################################

% Med e var KLD. Med e var FAD... destaques de min e max
% Depois por gênero pra cada um

% 1 - análise nas linhas (modelos)
% GVMGen tem maior variância nos gêneros, enquanto o nosso tem variância menor
% FAD no geral tem uma variância grande no geral
% 2 - na coluna
% KLD -> todos foram piores em estratégia. Médias e variâncias são similares, mas ainda assim somos melhores.
% FAD -> Plat, Puz & Sho são os piores, mas somos os melhores.  
% ImgBind Puz, Rac e Spo são piores

% #################################
% Figure: Heatmap for SOTA Comparison
% #################################

\begin{figure*}[t!]

\begin{minipage}[b]{1.0\linewidth}
\centering
\centerline{\includegraphics[width=1.0\linewidth]
  {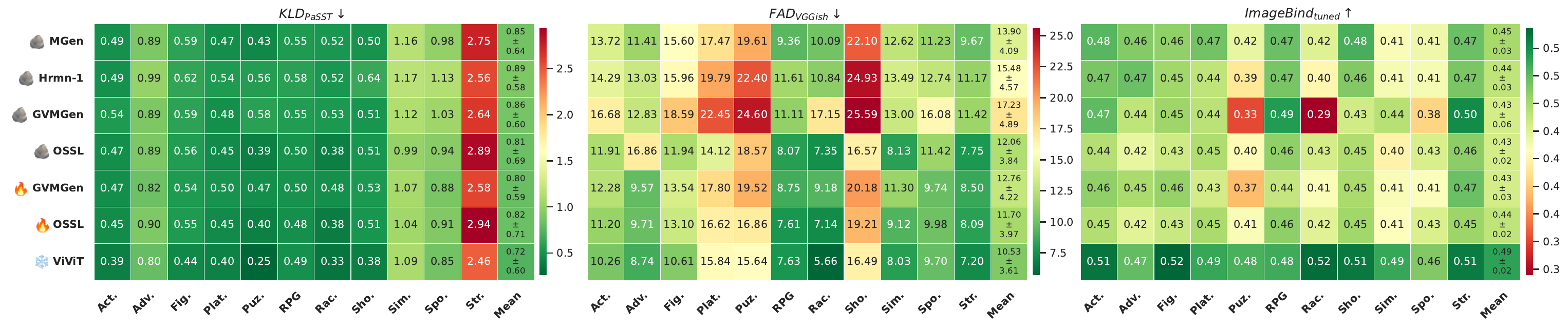}}
\end{minipage}

\caption{Objective performance comparison of the best proposed architecture against baseline models.}
\label{fig:heatmap}
\end{figure*}

\textbf{SOTA Comparison.} We compare \emoji{snowflake} ViViT, our best model, against one SOTA approach per category in Section \ref{sec:related_work}: Herrmann-1 ~\cite{herrmann1} for text as interface, OSSL~\cite{ossl} for text embedding as interface, and GVMGen~\cite{gvmgen} for direct mapping. For OSSL and GVMGen, we consider both their base (\emoji{stone}) versions, with their original pre-trained weights, and fine-tuned (\emoji{fire}) versions, whose adapters are trained from scratch on our dataset with the same hyperparameters as our models. For fine-tuned OSSL, all weights are frozen except the LoRA adapter~\cite{ossl}; for fine-tuned GVMGen, the ViT encoder is frozen, while the Feature Transformation module and MusicGen decoder are trained~\cite{gvmgen}. Herrmann-1 uses only frozen models, with descriptions generated as in~\cite{herrmann1}, except that we omit the speech module and replace GPT-4 with Qwen2.5-7B-Instruct. We also report MusicGen Base, which, together with OSSL, uses the same captions as our T5 model.

Fig.~\ref{fig:heatmap} shows one heatmap per objective metric, where rows correspond to models, columns to genres, and the last column to the mean and standard deviation across genres; greener cells indicate better scores. \emoji{snowflake} ViViT achieves the best mean on all metrics while using fewer parameters (approximately 1.93B, compared with 2.35B for OSSL and 2.22B for GVMGen), and it ranks first or tied in at least 9 of the 11 genres on every metric. Genre difficulty is consistent across models: Strategy yields the highest KLD, likely because its top-down map views are more ``open to interpretation,'' while Platform, Puzzle, and Shooter yield the highest FAD. 

Two comparisons isolate the effect of our design. First, OSSL uses the same frozen ViViT encoder but maps it into the T5 embedding space of a frozen decoder; our direct mapping with a trained decoder outperforms it on all metrics. Second, GVMGen, like our \emoji{snowflake} ViT model, pairs a frozen ViT with a trained decoder, but adds a Feature Transformation module; our simpler model again performs better. Moreover, fine-tuning substantially improves GVMGen, whose decoder is trained, but barely changes OSSL, whose decoder stays frozen, reinforcing that adapting the decoder is the key step. Finally, base GVMGen, trained on real-world videos, yields the worst FAD, even below MusicGen Base, highlighting the domain gap addressed by our dataset, while Herrmann-1, which translates descriptions into musical prompts with an LLM, underperforms MusicGen Base conditioned on raw captions, supporting our choice to omit this step.

\textbf{Listening Test.} Since~\emoji{snowflake} ViViT outperformed the other encoders on the objective metrics, we compared it against the fine-tuned OSSL and GVMGen in an online listening test, following the within-subjects methodology used to evaluate OSSL~\cite{ossl}. Each participant was randomly assigned two gameplay clips from a pool of 122 test clips. For each clip, participants first watched the original gameplay to become familiar with its visual dynamics and background music. They then watched the same clip three more times, each with background music generated by a different model in randomized order, and rated each generated soundtrack on five 5-point Likert scales: Q1) Video-Music Alignment, Q2) Game Genre Alignment, Q3) Audio Quality, Q4) Compositional Professionalism, and Q5) SNES Aesthetic Alignment. This procedure was then repeated for the second clip. We recruited participants via university mailing lists and public gaming-related Discord channels, allowing further snowball sampling. In total, $N=96$ participants completed the test.

Table~\ref{tab:listening_test} shows the average score of each method for each question. For each question, we ran a Kruskal-Wallis test across methods and, if significant, pairwise Wilcoxon rank-sum tests with Holm correction. The methods differ significantly ($p \leq 0.05$) only on Q1, Q2, and Q5. On video-music alignment (Q1) and game genre alignment (Q2), ViViT and OSSL perform comparably, and both significantly outperform GVMGen. On SNES aesthetic alignment (Q5), ViViT achieves the highest score, significantly outperforming GVMGen. Given that ViViT is nearly $18\%$ smaller than OSSL, these results suggest that frozen spatiotemporal representations are an effective choice for retro-game music generation.

% #################################
% Table: User Exp.
% #################################
\begin{table}[t]
 \centering
\begin{tabular}{llll}

\toprule

& \textbf{\emoji{fire} OSSL} & \textbf{\emoji{fire} GVMGen} & \textbf{\emoji{snowflake} ViViT}\\ 

\midrule

\textbf{Q1} $\uparrow$ & $2.82\pm1.14^{a}$ & $2.53\pm1.28^{b}$ & $\mathbf{2.87}\pm1.21^{a}$\\ 

\textbf{Q2} $\uparrow$ & $2.99\pm1.15^{a}$ & $2.76\pm1.23^{b}$ & $\mathbf{3.06}\pm1.17^{a}$\\ 

\textbf{Q3} $\uparrow$ & $\mathbf{3.38}\pm1.09$ & $3.22\pm1.16$ & $3.26\pm1.11$\\ 

\textbf{Q4} $\uparrow$ & $\mathbf{3.05}\pm1.14$ & $2.88\pm1.22$ & $2.84\pm1.23$\\ 

\textbf{Q5} $\uparrow$ & $3.29\pm1.13^{ab}$ & $3.06\pm1.25^{a}$ & $\mathbf{3.40}\pm1.15^{b}$\\ 
\bottomrule
\end{tabular}
\caption{Average scores (in 1–5) for each metric in our listening test. % The metrics are: Q1) Video-Music Alignment, Q2) Game Genre Alignment, Q3) Audio Quality, Q4) Compositional Professionalism, and Q5) SNES Aesthetic Alignment. 
Values with shared letter superscripts$^{ab}$ in a given row have p-value $> 0.05$ in pairwise post-hoc comparisons.} %A value in bold is the best score for a particular metric (row).}
\label{tab:listening_test}
\end{table}

% #################################
% Conclusion
% #################################

\section{Conclusion}

% #TODO: - (Conclusion, L515-517) "suggesting that the features of large-scale pre-trained models transfer to the video game domain." -> This claim is problematic. The fact that fine-tuning doesn't help doesn't suggest that the pretrained models generalize to the target domain. In fact, it's more often the other case -- neither the pretrained models generalize to the target domain, nor does finetuning help. I don't think this claim itself is wrong though, as can be seen by the good performance of these models. Please support this claim with other evidences.

% #TODO: - (Conclusion, L518) "state-of-the-art performance" -> Please scope this claim properly -- on what dataset under what setup?

We introduced an SNES video-to-music dataset and used it to compare visual encoding strategies for autoregressive video-to-music generation. Among T5, ViT, and ViViT encoders, ViViT achieved the best overall results, and frozen encoders matched or outperformed fine-tuned ones, suggesting that real-world pre-trained features transfer to rendered game graphics when only the decoder is adapted. Our simple direct-mapping model also achieved the best objective results against SOTA baselines and performed comparably to OSSL in a listening study, with up to 18\% fewer parameters. 
% Future work includes improving audio mapping accuracy and extending our pipeline to other consoles.

% While these are important results for the video-to-music literature, current models have yet to reach human-level performance. Consequently, future work will explore integrating our SNES dataset with the NES-VMDB and other video game corpora to investigate how these models generalize across different console generations. Additionally, by incorporating more recent platforms featuring realistic graphics and high-fidelity audio, we aim to investigate how modern video games can contribute to generating music for broader applications, such as films or music videos.

% #################################
% Acknowledgements
% #################################
% \section{Acknowledgments}
% We would like to thank the Universidade Federal de Viçosa (UFV) for access to the DGX GPU Cluster. In particular, we are grateful for the long-term technical support given by Eliab Venancio, head of the Scientific Development Support Service at the Diretoria de Tecnologia de Informação (DTI).

% References should be produced using the bibtex program from suitable
% BiBTeX files (here: strings, refs, manuals). The IEEEbib.bst bibliography
% style file from IEEE produces unsorted bibliography list.
% -------------------------------------------------------------------------
\bibliographystyle{IEEEbib}
\bibliography{refs}

\end{document}